\documentclass[conference]{IEEEtran}
\IEEEoverridecommandlockouts
\usepackage{fancybox}
\usepackage[most]{tcolorbox}
\usepackage{url}
\usepackage[numbers,sort]{natbib}
\usepackage{amsmath,amssymb,amsfonts}
\usepackage[ruled,linesnumbered, noend, noline]{algorithm2e}
\usepackage{varwidth}
\usepackage{graphicx}
\usepackage{textcomp}
\usepackage{enumitem}
\usepackage{mathtools}
\usepackage[table]{xcolor}
\usepackage[hidelinks]{hyperref}
\usepackage[capitalise, nameinlink, noabbrev]{cleveref}
\usepackage{listings}
\usepackage{relsize}
\usepackage{siunitx}
\usepackage{xspace}
\usepackage{balance}
\usepackage{blindtext}
\usepackage[T1]{fontenc}
\usepackage{tabularx}
\usepackage{booktabs}
\usepackage{multirow}
\usepackage{enumitem}

\definecolor{lightcyan}{rgb}{0.88, 1.0, 1.0}
\definecolor{mistyrose}{rgb}{1.0, 0.89, 0.88}
\definecolor{lightyellow}{rgb}{1.0, 1.0, 0.88}
\definecolor{lightgray}{rgb}{0.83, 0.83, 0.83}

\definecolor{eclipseBlue}{RGB}{42,0.0,255}
\definecolor{eclipseGreen}{RGB}{63,127,95}
\definecolor{eclipsePurple}{RGB}{127,0,85}
\definecolor{lightblue}{RGB}{58, 125, 201}
\definecolor{stringgreen}{RGB}{35, 107, 51}
\definecolor{verylightgray}{RGB}{248, 248, 248}

\title{Aging of Prompt Engineering Techniques Across LLM Versions}

\author{Anastasiia Rudyk \\
Software Engineering \\
University of Rostock \\
{anastasiia.rudyk@uni-rostock.de}
\and
Julian Oertel \\
Software Engineering \\
University of Rostock \\
{julian.oertel@uni-rostock.de}
\and
Regina Hebig \\
Software Engineering \\
University of Rostock \\
{regina.hebig@uni-rostock.de}
\thanks{Accepted for publication at the 42nd IEEE International Conference
  on Software Maintenance and Evolution (ICSME 2026), Benevento, Italy.
  This is the author's accepted version. \copyright~2026 IEEE. Personal use
  of this material is permitted. Permission from IEEE must be obtained for
  all other uses, in any current or future media, including
  reprinting/republishing this material for advertising or promotional
  purposes, creating new collective works, for resale or redistribution to
  servers or lists, or reuse of any copyrighted component of this work in
  other works.}
}

\makeatletter
\newcounter{RQCounter}
\newenvironment{ResearchQuestion}{%
\par\nointerlineskip%
\refstepcounter{RQCounter}%
\protected@edef\@currentlabelname{RQ\theRQCounter}%
\protected@edef\@currentlabel{RQ\theRQCounter}%
\vspace{0.15cm}\noindent%
\begin{minipage}[b]{\columnwidth}%
\textbf{RQ\theRQCounter: }}{%
  \end{minipage}%
}%
\makeatother
\begin{document}

\maketitle

\begin{abstract}
Prompt engineering and prompt engineering techniques (PETs) have become an integral part of software engineering for AI systems.
However, new LLMs are released frequently and it remains unclear how the effectiveness of prompt engineering techniques changes across successive generations of Large Language Models (LLMs).
To this end, we conduct a partial replication of the study by Khojah et al. (2025).
We evaluate five techniques -- Zero-Shot, Few-Shot, Chain-of-Thought (CoT), Contrastive Chain-of-Thought (CCoT), and an adapted version of Program-of-Thought (PoT) -- on six instruction-tuned models grouped into three version pairs: GPT-3.5-Turbo → GPT-4o, Qwen2 7B Instruct → Qwen2.5 7B Instruct, and Mistral-7B-Instruct → Mistral-Large. We use a cleaned subset of the CodePromptEval dataset with 218 context-rich Python functions and 19,620 total generations assessed via pass@k-based functional correctness to evaluate model pairs on function-level code generation tasks.
We show that prompt engineering ``ages'' in a model-family-specific way: Newer GPT models exhibit diminishing or even negative marginal gains from structured prompting, suggesting that instruction-following and reasoning scaffolds are increasingly internalized, whereas Qwen models continue to benefit substantially from Few-Shot and CCoT. Mistral models show mixed behavior with persistent gains from CCoT but attenuated benefits from CoT and PoT. 
Our results imply that effective prompting strategies must be adapted per model family and generation rather than transferred unchanged. This motivates future work on adaptive, model-aware prompting and broader, multi-dimensional code quality evaluation. 
\end{abstract}
\section{Introduction}
%
The increasing use of LLMs in software engineering, particularly 
for automated code generation, necessitates effective input strategies to guarantee high-quality 
outputs~\cite{shinPromptEngineeringFineTuning2025, bruniBenchmarkingPromptEngineering2025a, fengPromptingAllYou2024}. Creating strategic textual inputs, or prompts, to direct LLMs and optimize their intended performance without changing model parameters is known as prompt engineering~\cite{chenUnleashingPotentialPrompt2025, 10.1145/3560815, sahooSystematicSurveyPrompt2025}.

The rapid progression of artificial intelligence, particularly in the development of LLMs, has resulted in frequent releases of new model generations that exhibit improved reasoning abilities, scalability, and domain adaptability~\cite{openaiModelsGPT4o, Gemini31Pro, yangQwen3TechnicalReport2025}. This accelerating pace has fostered extensive research comparing the performance of distinct LLM architectures across tasks such as code completion, text summarization, and reasoning~\cite{naveedComprehensiveOverviewLarge2025, dingReasoningPlanningLarge2024, wangAdvancedLanguageModels2025, khojahImpactPromptProgramming2025}. 

Concurrently, the refinement of prompt engineering has gained significant scholarly attention. Studies have examined how various prompting strategies influence code quality and performance metrics, including accuracy, maintainability, and reliability~\cite{dellaportaPromptPatternsAffect2025, wangAdvancedLanguageModels2025}. However, despite these advances, there remains a notable gap in systematically analyzing how the effectiveness of prompt engineering techniques evolves across successive versions of different LLMs. Addressing this gap is vital to understanding whether improvements in model design inherently reduce prompt sensitivity or whether prompt optimization remains critical for maintaining output quality.

Khojah et al. created CodePromptEval, a dataset of 7,072 prompts using a full factorial design to test five PETs (few-shot, persona, chain-of-thought, function signatures, and package lists) across project-level code generation tasks~\cite{khojahImpactPromptProgramming2025}. They examined GPT-4o, Llama3-70B-Instruct, and Mistral-22B-Instruct (also GPT-3.5-Turbo and Llama2-7B-Instruct without detailed reporting) solving real open-source function tasks, measuring impacts on correctness, similarity, and code quality. Their results showed that function signatures and few-shot examples boost functional correctness reliably, but stacking all five techniques fails to deliver consistent gains and risks adding code smells.

In this paper, we investigate the evolution of prompting techniques across successive LLM generations, driven by the challenge that LLM performance 
often remains unstable and highly sensitive to input format in software engineering tasks, necessitating careful prompt engineering~\cite{jiBenchmarkingExplainingLarge2023, khojahImpactPromptProgramming2025}. This study systematically compares the effectiveness of fundamental prompting techniques, such as Zero-shot and Few-shot approaches~\cite{radford2019language, brownLanguageModelsAre2020a}, alongside advanced reasoning-based methods, including CoT~\cite{NEURIPS2022_9d560961}, CCoT~\cite{chiaContrastiveChainofThoughtPrompting2023}, and an adapted version of PoT~\cite{chenProgramThoughtsPrompting2023}.
With this we aim to answer the following research questions.
\begin{ResearchQuestion}\label{rq1}
\textbf{How does the performance of different prompting strategies vary across successive versions of LLMs in function-level code generation tasks?}
\end{ResearchQuestion}

We use a modified version of the CodePromptEval dataset, to benchmark prompting effectiveness across proprietary models (GPT-3.5-Turbo, GPT-4o) and open-source models (Qwen2/Qwen2.5 7B Instruct, Mistral-7B-Instruct/Large). We replicate the original study with a few notable distinctions: (1) We substitute the original PETs ``signature'' and ``persona'' with CCoT and the adapted PoT to not only replicate, but extend the results to other PETs; (2) We do not adhere to the full factorial design as we wanted to focus on individual effects only and (3) except for GPT-4o and GPT-3.5-Turbo, we utilize different LLMs due to cost constraints and our goal of comparing subsequent model versions.


\section{Related Work}
\label{sec:related_work}

This section positions the present study within the broader academic landscape by reviewing prior research through a systematic funnel approach. The discussion progresses from foundational advancements in LLM and their evolution, through evaluation methodologies and benchmarking practices. 

\subsection{Benchmarking and Evaluation Methodologies}
\label{sec:benchmarking}

Rigorous evaluation frameworks are essential for assessing LLM capabilities in code generation. Multiple benchmarks have been developed, each emphasizing different aspects of code synthesis and addressing limitations of prior evaluation approaches.

\subsubsection{Established Benchmarks and Pass@k Metrics}

HumanEval, released with Codex, includes 164 hand-crafted programming problems featuring function signatures, docstrings, and unit tests to assess functional correctness~\cite{chenEvaluatingLargeLanguage2021}. The study utilized the pass@k metric, originally proposed by Kulal et al. (2019)~\cite{NEURIPS2019_7298332f}, but implemented a novel unbiased estimator to reduce sampling variance during probabilistic evaluation. This approach acknowledges that multiple samples from the same model often succeed where single attempts fail on challenging prompts~\cite{chenEvaluatingLargeLanguage2021}, and has become the de facto standard for evaluating code generation models.

MBPP extends this evaluation with 974 crowd-sourced problems emphasizing fundamental programming constructs~\cite{austin2021mbpp}.

\subsubsection{Context-Dependent Function Generation}

Subsequent benchmarks have addressed limitations in these foundational datasets by emphasizing more realistic code generation scenarios. CoderEval introduced 460 tasks (230 Python, 230 Java) emphasizing non-standalone functions that invoke project-specific API, reflecting realistic development scenarios where over 70\% of functions depend on external context~\cite{yuCoderEvalBenchmarkPragmatic2024}. This benchmark revealed substantial performance gaps: models achieving high accuracy on standalone functions showed markedly reduced effectiveness on context-dependent generation, highlighting that evaluation on isolated functions may overestimate practical utility.

\subsection{Empirical Studies of Prompt Engineering Effectiveness}
\label{sec:prompt_effectiveness}

Numerous empirical studies have investigated how PET influence code generation performance, revealing complex patterns of effectiveness that vary across models, tasks, and evaluation metrics.

\subsubsection{Model-Dependent Effectiveness}

Murr et al. investigated how levels of prompt specificity impact code generation performance~\cite{murrTestingLLMsCode2023}. After testing four models with over 100 problems and various prompt structures, the authors noted that outcomes fluctuated markedly depending on both the specific language model and the prompting method employed. Crucially, optimal prompting strategies differ substantially by model capability, with more capable models requiring less explicit guidance. This finding suggests that prompt engineering is not a universal solution but rather a model-dependent practice requiring calibration to specific model characteristics.

Liu et al. similarly assessed ChatGPT's code generation quality, revealing competent functional output but deficiencies in efficiency, security, and standalone usability—often requiring human intervention despite the 'no finger-lifting' promise~\cite{10507163}.

Della Porta et al. investigated whether prompt patterns (Zero-Shot, Few-Shot, CoT and Personas) significantly affect code quality metrics including maintainability, security, and reliability~\cite{dellaportaPromptPatternsAffect2025}. In their analysis of 7,583 code files from the Dev-GPT dataset via Kruskal-Wallis tests, the authors detected few quality problems overall and no notable variations between prompt patterns. These findings suggest that, within the scope of their dataset and measurement approach, variations in prompt structure were not associated with measurable differences in the assessed structural quality attributes. Accordingly, prompt format alone may not be a decisive factor for maintainability- and security-related metrics in ChatGPT-assisted code generation, at least under the conditions examined in their study.

\subsubsection{Diminishing Returns in Advanced Models}

Recent empirical research has investigated whether prompt engineering benefits persist as models advance. Wang et al. conducted extensive evaluations across code generation, translation, and repair tasks, comparing GPT-4o and reasoning-native o1-mini~\cite{wangAdvancedLanguageModels2025}. Their findings revealed that non-reasoning models still benefit substantially from structured prompts, while reasoning models often achieve comparable or superior performance with simple zero-shot instructions. This pattern stems from a fundamental difference: reasoning models internalize complex reasoning scaffolds natively, reducing dependence on explicit prompt structures.

The quantitative evidence supporting these observations is substantial. Analysis of problems requiring extended reasoning chains reveals a 16.67\% performance advantage for o1-mini over GPT-4o when the chain-of-thought exceeds five steps, compared to only 2.89\% for simpler problems~\cite{wangAdvancedLanguageModels2025}. This performance gap suggests that the advantages of reasoning-native models become substantially more pronounced as the logical complexity of the programming task increases. 


\subsubsection{Causality-Based Analysis and Linguistic Features}

Ji et al. employed a causality-centric approach to analyze how linguistic features in prompts -- such as formal versus fluent phrasing -- affect code quality across GPT-Neo, GPT-3.5, and GPT-4~\cite{jiBenchmarkingExplainingLarge2023}. Their causal graph-based representation revealed trade-offs where specific keywords significantly alter code patterns, suggesting that model sensitivity to prompt design is version-dependent. This work offered practical guidance for refining prompts to enhance LLM-generated code quality, while revealing that prompt effectiveness varies across models and demands testing for each version.

Wang et al. proposed PET-Select, a complexity-based selection mechanism that classifies queries and automatically chooses appropriate PET~\cite{wangSelectionPromptEngineering2024}. By using code complexity as a proxy for task difficulty, PET-Select achieved 1.9\% improvement in pass@1 accuracy while reducing token usage by 74.8\%. They demonstrate that adaptive prompt selection outperforms static approaches and suggest future work should focus on refining the selection model and extending its application to broader domains.





\section{Description of the Original Study}
In line with best practices outlined by Carver~\cite{carver2010towards}, we will describe the original study in this section.
\subsection{Data Set: CodePromptEval}
With their study Khojah et al. introduce the data set CodePromptEval which is the foundation of their subsequently presented work~\cite{khojahImpactPromptProgramming2025}.
CodePromptEval is adapted from CoderEval, which consists of 230 Python functions and 230 Java functions extracted from popular GitHub repositories. 
Each code generation task then consists of an original docstring or comment, a function signature, source code, and associated tests. A difficulty is assigned by considering the code dependencies, e.g., implementing a function that is self-contained is regarded as easier than implementing a function that re-uses code from a different class.

For CodePromptEval Khojah et al. select 221 (thereby excluding tasks with failing ground truth) of the Python tasks and create 32 prompts for each combination of their chosen PETs (cf. \Cref{subsec:original_method}) as well as a zero-shot prompt used as a baseline.

Afterwards, they use GPT-4o, Llama3-70B-Instruct, Mistral-Small-Instruct-2409 (22B), GPT-3.5-turbo, and Llama2-7B-Instruct to generate 3 solutions to each prompt using a temperature of \(0.2\). 

\subsection{Research Questions}
With their data they answer the following research questions:

\noindent
RQ1: How do different LLMs perform on CodePromptEval?

\noindent
RQ2: To what extent do different prompting techniques (and combinations of them) impact the code generation of LLMs?

They split their second research question further and aim to investigate the impact on code correctness, code similarity to human-written code, and code quality.

\subsection{Prompting Techniques}
\label{subsec:original_method}
\begin{figure}
    \centering
    \includegraphics[width=\linewidth]{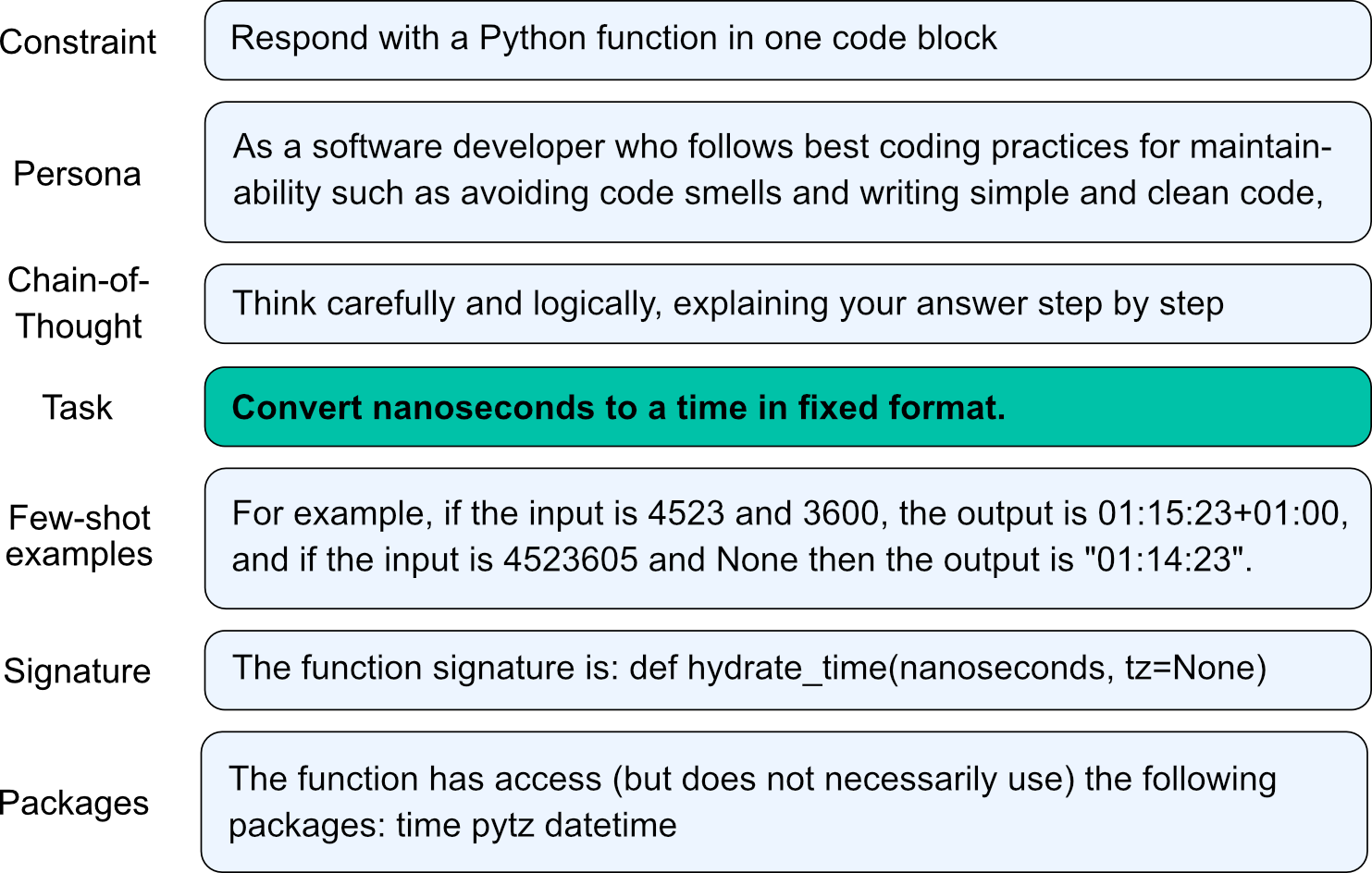}
    \caption{Example prompt from Khojah et al.~\cite{khojahImpactPromptProgramming2025}}.
    \label{fig:original_prompt}
\end{figure}

Khojah et al. evaluate 6 different PETs with an example illustrated in \Cref{fig:original_prompt}. Namely, they look into zero-shot, few-shot, persona and chain-of-thought prompting as well as techniques they call ``signature'' and ``packages''.
They define zero-shot as their baseline for evaluation purposes as it cannot be combined with other prompting techniques. In this case they provide only the constraint shown in \Cref{fig:original_prompt} to ease automation and extract the code automatically later on. This constraint is also used in every other prompt, regardless of the otherwise used techniques. 
Few-shot prompting is implemented by providing two input-output pairs to illustrate the desired behavior. Persona is implemented by instructing the LLM to act as a software developers who adheres to best practices. 
For Chain-of-Thought, they use a zero-shot chain-of-thought as proposed by Kojima et al.~\cite{Kojima} that simply instructs the LLM to think step by step. 
In ``Signature'' and ``packages'', the LLM is provided with additional information: either the function signature with the expected arguments or information about packages that could be used as they are accessible in the function.

\subsection{Analysis Procedure}
\label{subsec:original_procedure}
To answer their research questions, they extract metrics from their generated functions. They reuse the original tests provided in CoderEval~\cite{yuCoderEvalBenchmarkPragmatic2024} to test for correctness. Based on this, they calculate the pass@k metric. For code similarity to the human-written baseline they use the CrystalBLEU metric. For code quality they use Pylint to detect code smells in the generated code and they calculate McCabe complexity and cognitive complexity. 

Their subsequent analysis is focused on descriptive metrics for RQ1. For RQ2 and the sub-research questions they additionally use regression models and statistical tests. 

\subsection{Results}
For RQ1, Khojah et al. observe that GPT-4o outperforms the other studied LLMs slightly while all LLMs achieve a similar performance. Depending on the task difficulty all LLMs could solve between 31\% and 90\% of tasks~\cite{khojahImpactPromptProgramming2025}. 

For RQ2 they find that signature and few-shot had the clearest impact on correctness, finding no statistically significant impact for the other PETs. However, the general differences in correctness are low and providing more information sometimes worsens the results.
CoT, persona, and package decreased the number of code smells but also decreased the general correctness.
The similarity to the human-written baseline code increased with signature and persona. Few-shot mostly decreased similarity. The code complexity of generated solutions was generally similar or lower compared to the human-written code~\cite{khojahImpactPromptProgramming2025}. 



\section{Replication Study Design}
In this section, we outline the procedure of this study and highlight differences to the original work of Khojah et al.~\cite{khojahImpactPromptProgramming2025}.
An overview that shows the extent of our partial replication is given in \Cref{tab:differences} and explained in more detail in the following.
\begin{table}[t]
\caption{Overview of aspects recreated and changed in this study compared to Khojah et al.\cite{khojahImpactPromptProgramming2025}}
    \centering
    \begin{tabular}{p{1.0cm}>{\raggedright\arraybackslash}p{3cm}>{\raggedright\arraybackslash}p{3.1cm}}
    \toprule
         Aspect & Khojah et al. &  This Study\\
         \midrule
       Goal & Combining PETs in a full factorial design to find suitable combinations & Examining changes in PETs across model versions\\
       \midrule
       PETs & Zero-shot, Few-shot, CoT, Persona, Packages Signature & Zero-shot Few-shot, CoT, Contrastive-CoT, Adapted Program-of-Thought\\
       \midrule
       LLMs & GPT-4o, Llama3-70B-Instruct, Mistral-Small-Instruct-2409 (22B) &  GPT-3.5-Turbo, GPT-4o; Qwen2 7B Instruct, Qwen2.5 7B
Instruct; Mistral-7B-Instruct, Mistral-Large-2407 (123B)\\
       \midrule
       Task Data Set & CodePromptEval& CodePromptEval (3 tasks excluded)\\
       \midrule
       Compared Metrics & Pass@k, CrystalBLEU, Cognitive and cyclomatic complexity, code smells& Pass@k\\
       
       \bottomrule
    \end{tabular}
    
    \label{tab:differences}
\end{table}
\subsection{Prompt Data Set}
\begin{figure}
    \centering
   \includegraphics[width=0.6\linewidth]{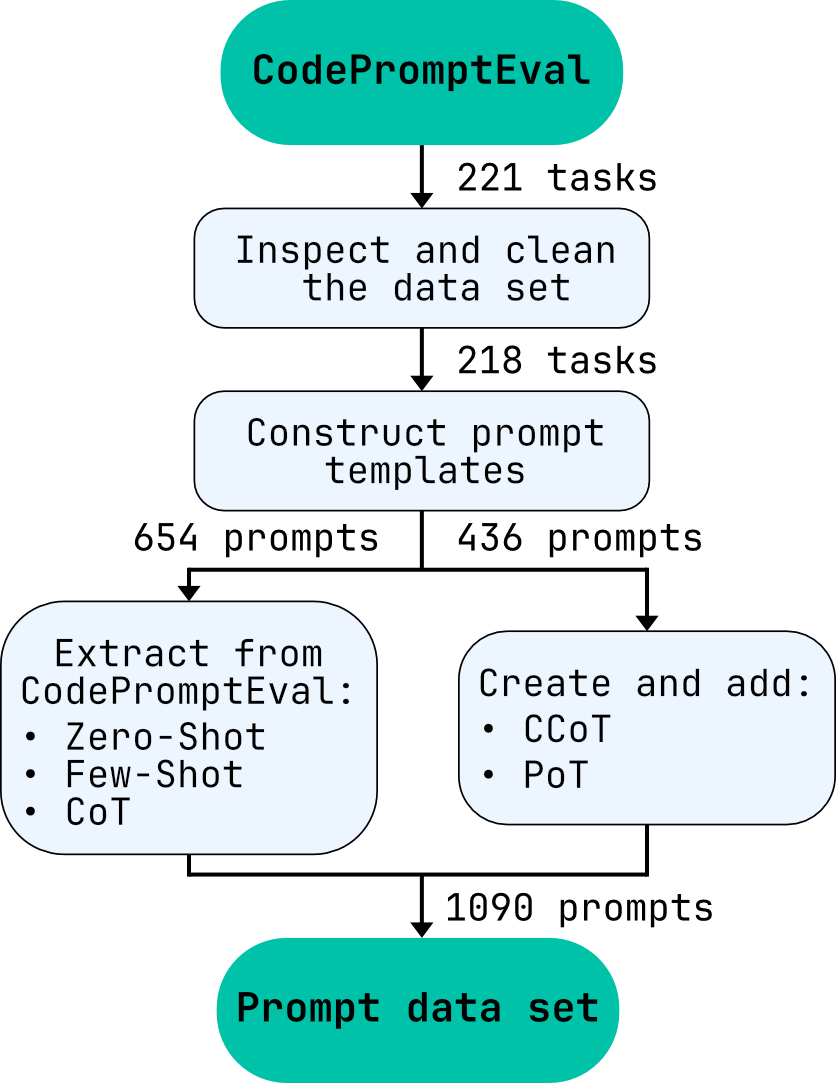}
    \caption{Creation of prompt data set used for this study}
    \label{fig:procedure}
\end{figure}

 \Cref{fig:procedure} outlines our general procedure in obtaining the data set.
\subsubsection{Cleaning of Original Data Set}
CodePromptEval consists of 221 tasks. In preparation for our study, we re-examined the proposed tasks and found that three tasks did not adhere to the format required by few-shot prompting. These tasks did not define input and output in their example or their example included wrong information. We thus excluded the three tasks and all associated prompts, leaving us with 218 tasks. 
\begin{table}[t]
\centering
\caption{Prompt templates used in the replication}
\label{tab:prompt_templates_table}
\scriptsize
\begin{tabular}{p{1.0cm}p{7.0cm}}
\toprule
\textbf{Name} & \textbf{Template} \\
\midrule
Zero-Shot & Respond with a Python function in one code block. \newline\textbf{\texttt{\{Coding task\}}}. \\
\midrule
Few-Shot & Respond with a Python function in one code block. \newline\textbf{\texttt{\{Coding task\}}}. \newline\textbf{\texttt{\{Two examples\}}}.\\
\midrule
Chain-of-Thought & Respond with a Python function in one code block. Think carefully and logically, explaining your answer step by step.\newline\textbf{\texttt{\{Coding task\}}}. \\
\midrule
Contrastive Chain-of-Thought & Respond with a Python function in one code block. Think carefully and logically, explaining your answer step by step.\newline\textbf{\texttt{\{Coding task\}}}.\newline\textbf{\texttt{\{Two positive examples\}}}. \newline Here is an example of a wrong explanation: \textbf{\texttt{\{Wrong example 1\}}}. \newline Here is another example of a wrong explanation: \textbf{\texttt{\{Wrong example 2\}}}. \\
\midrule
Adapted Program-of-Thought & Respond with a Python function in one code block. Think carefully and logically, explaining your answer step by step. Use code to express all reasoning steps. \newline\textbf{\texttt{\{Coding task\}}}. \\
\bottomrule
\end{tabular}
\end{table}
\subsubsection{Prompt Engineering Techniques}
We employ five PETs: three adapted from CodePromptEval (Zero-Shot, Few-Shot, CoT) and the additional techniques CCoT~\cite{chiaContrastiveChainofThoughtPrompting2023} and an adapted version of PoT.
The template for each PET is outlined in \Cref{tab:prompt_templates_table}. Zero-shot, Few-shot and CoT stayed true to the original study. 

For CCoT, the positive examples were derived from a combination of the Few-Shot and Chain-of-Thought techniques used in the CodePromptEval dataset. Following Chia et al.'s original methodology, the negative examples were created by generating 2–3 incorrect demonstrations. This was achieved by using the open-source entity recognition model\footnote{ \url{https://spacy.io/models/en\#en_core_web_trf}} to identify the core objects within the correct rationale and randomly shuffling the position of these objects within the reasoning chain.
The core objects considered were key variables, numbers, function names, and constants (None, True, False).
This method was found to be the most effective way to construct generalized negative examples~\cite{chiaContrastiveChainofThoughtPrompting2023}.

The original PoT technique is used to address logical and numerical reasoning limitations using executable code and therefore effectively transfers these tasks to code generation tasks~\cite{chenProgramThoughtsPrompting2023}. As we are already generating code, we instead opted to use a programmatic step-by-step thinking to foster a reasoning process that is closer related to the final output. To achieve this, we extended the CoT prompt by instructing the LLM to ``use code to express all reasoning steps''.

Overall, we thus extracted 654 from CodePromptEval and created an additional 436 which add up to a total of 1090 prompts. 

\subsection{Generated Data Set}
\begin{figure}
    \centering
    \includegraphics[width=0.65\linewidth]{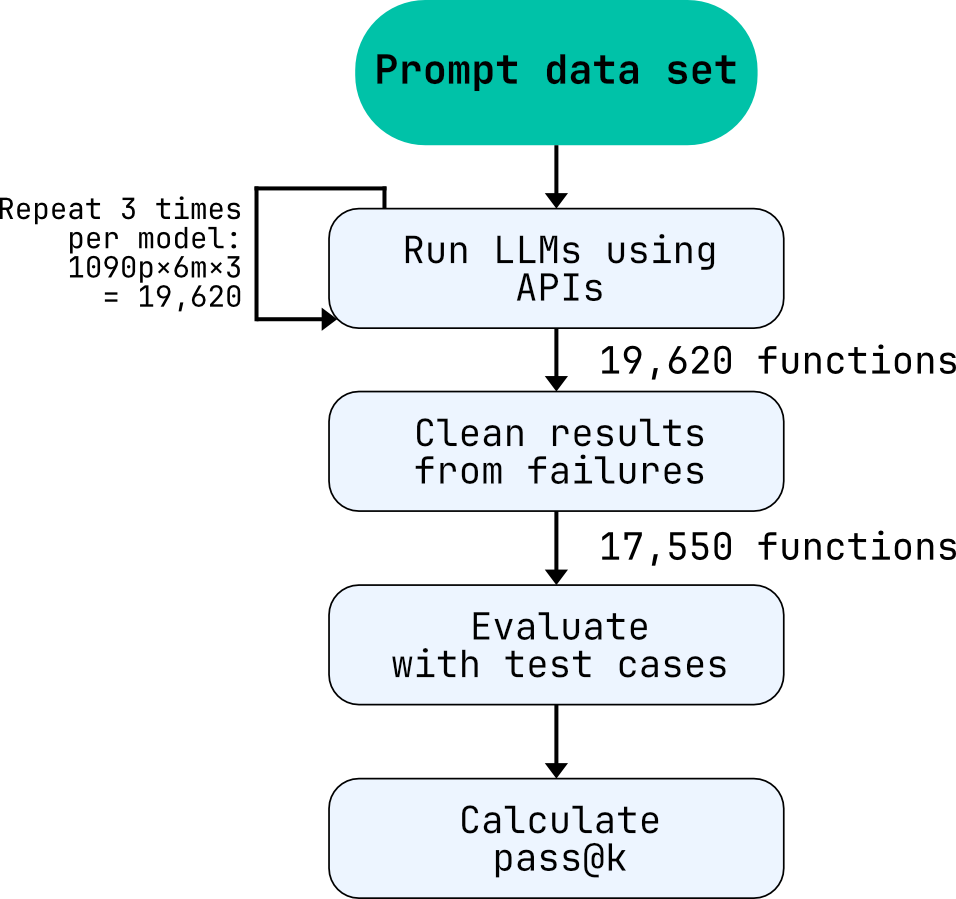}
    \caption{Procedure for generating, extracting and evaluating the code}
    \label{fig:generation_procedure}
\end{figure}

In this section, we present the LLMs and procedure used to generate and evaluate the code corresponding to the prompts. \Cref{fig:generation_procedure} outlines the general procedure. 

\subsubsection{LLM Selection}
Model selection was guided by several criteria.
\paragraph{Training cut-off date} The CodePromptEval dataset is based on the CoderEval dataset, which was published in February 2024~\cite{yuCoderEvalBenchmarkPragmatic2024}. Consequently, this study considers only model versions with training data cut-off dates prior to this release. As CoderEval uses open-source code from GitHub, we cannot rule out data leakage completely. However, Yu et al. already tried to mitigate such risks during the construction of the data set.

GPT-3.5 Turbo has a training data cut-off date of September 2021, as reported in the official model documentation under Microsoft’s Azure AI Foundry catalog\footnote{\url{https://learn.microsoft.com/en-us/azure/ai-foundry/foundry-models/concepts/models-sold-directly-by-azure}}. The GPT-4o model, released in 2024, features a later cut-off in October 2023, according to the same source. 

For the Qwen series, a statement from a Qwen3 collaborator on the official GitHub repository indicates that the model’s training data, covering both pre-training and post-training stages, includes material up to the end of 2023\footnote{\url{https://github.com/QwenLM/Qwen3/issues/525##issuecomment-2159944330}}. As Qwen3 is the direct successor to Qwen2.5~\cite{yangQwen3TechnicalReport2025}, this time frame has been widely adopted by the research community as the effective data cut-off for both Qwen2 and Qwen2.5\footnote{\url{https://github.com/HaoooWang/llm-knowledge-cutoff-dates}}. 

Likewise, within the Mistral family, publicly available configuration files document that Mistral-Large employs a training cut-off of October 2023\footnote{\url{https://www.allmo.ai/articles/list-of-large-language-model-cut-off-dates}}. The Mistral 7B Instruct variant’s knowledge cut-off date has been identified as December 2023 in multiple community discussions and blogs\footnote{\url{https://huggingface.co/mistralai/Mistral-7B-Instruct-v0.2/discussions/69}}.

\paragraph{Suitability for code generation}The selection of models was based on their demonstrated effectiveness in code generation as validated by authoritative benchmarks and official documentation. GPT-3.5 Turbo\footnote{\url{https://platform.openai.com/docs/models/gpt-3.5-turbo}} and GPT-4o\footnote{\url{https://openai.com/index/hello-gpt-4o/}} are marketed by OpenAI specifically for coding tasks, with multiple benchmark studies confirming their high performance in code generation and reasoning~\cite{wangSelectionPromptEngineering2024}. 

Qwen2 7B Instruct\footnote{\url{https://huggingface.co/Qwen/Qwen2-7B-Instruct}} and Qwen2.5 7B Instruct\footnote{\url{https://huggingface.co/Qwen/Qwen2.5-7B-Instruct}} have documented architectural optimizations for code generation and long-context reasoning, validated by technical documentation and benchmark results supporting their proficiency in code tasks. 

Mistral-7B-Instruct\footnote{\url{https://mistral.ai/news/announcing-mistral-7b}} and Mistral-Large 2\footnote{\url{https://mistral.ai/news/mistral-large-2407}} are noted in technical model releases for their instruction tuning aimed at coding scenarios, with reported competitive performance in instruction-following and code synthesis. Thus, inclusion criteria were formulated based on marketing positioning and rigorous benchmark evidence, allowing reproducibility and transparency in model selection for code generation studies.
\paragraph{Version traceability} Selected models should have clearly traceable successive versions, allowing for meaningful comparison across updates. The criteria for considering models as successive versions include sharing the same core architecture, consistent task orientation such as instruction tuning, and incremental improvements through updates in training data, hyperparameters, or scale. 

Based on these, GPT-3.5-Turbo and GPT-4o are successive iterations within OpenAI’s transformer-based models, with GPT-4o offering enhanced reasoning and efficiency. Similarly, Qwen2 7B Instruct and Qwen2.5 7B Instruct belong to the same architectural family with refined instruction tuning in the latter. Likewise, the Mistral-7B-Instruct and Mistral-Large models are scaled versions sharing the foundational architecture and tuning goals, supporting coherent comparisons across versions.
\subsubsection{Generation}
GPT-3.5-Turbo, GPT-4o, Qwen2.5 7B Instruct, Mistral-7B-Instruct and Mistral-Large were accessed via the OpenRouter API\footnote{\url{https://openrouter.ai}}, while Qwen2 7B Instruct was accessed via Replicate API\footnote{\url{https://replicate.com}}. For the Mistral-7B-Instruct model, all prompts were wrapped using the special \texttt{[INST]} and \texttt{[/INST]} tokens, as recommended in the “Instruction format” section of the official model card for Mistral-7B-Instruct\footnote{See “Instruction format” in the Mistral-7B-Instruct model card at \url{https://huggingface.co/mistralai/Mistral-7B-Instruct-v0.1}}. In line with the original study, we set a temperature of 0.2 for all models~\cite{khojahImpactPromptProgramming2025}. We additionally apply a maximum token limit of 1024 per generation attempt. 

Based on the 1090 prompts, this resulted in 19,620 responses over three runs.

\subsubsection{Code Extraction}
We automatically extract code of the responses by exploiting markdown formatting to identify relevant code. In cases where this failed, we tried to extract the code manually. 

For 23 prompts, we had problems in at least one of the generation attempts. Reasons for this were mostly responses without any code, but also generation attempts that were cut-off due to the token limit and generation attempts that resulted in endlessly repeated output. 

We opted to exclude all generation attempts for any of the 23 prompts, reducing our final data set of functions to evaluate to 17,550 and our effective number of tasks to 195.
\subsubsection{Code Evaluation}
For the evaluation of the generated code, we adopted the original approaches, preserving the original CoderEval Docker environments~\cite{yuCoderEvalBenchmarkPragmatic2024} and utilizing the CodePromptEval evaluation methodology~\cite{khojahImpactPromptProgramming2025}.

\subsection{Analysis Procedure}
As we are primarily interested in the correctness of code, we follow the original procedure outlined in \Cref{subsec:original_procedure} and focus on descriptive statistics focused on the pass@k metric.

\section{Results}
\label{chap:results}

This section presents the empirical findings from evaluating the five PETs across LLM versions.

\subsection{Comparison to Original Results}
We altered several components of the original study in order to answer our research question. However,
as we also included GPT-4o and re-used some of the PETs we can directly compare some of our results to those of Khojah et al.~\cite{khojahImpactPromptProgramming2025}.
\Cref{tab:result_comparison} highlights this comparison. To extract this data, we revisited the original replication package. As we excluded three tasks beforehand and another 23 tasks due to failure to extract code from the responses we also present the original results where we excluded these 23 tasks as well.
Overall, we observe consistently lower pass@1 rates for our replication with an average of 6.5 percentage point difference compared to the original results for the same 195 tasks. With an 8.0 percentage point difference, the largest reduction in pass@1 rates occurs for the few-shot PET. 

A comparison of the original data set against the reduced original data set yields an average difference of 0.07 percentage points with a maximum difference of 0.2 percentage points for the zero-shot PET.

\begin{table}[t]
    \centering
    \small
    \caption{Comparison between the original results by Khojah et al. and our replication for pass@1 results averaged over 3 runs.}
    \begin{tabular}{lrrr}
    
    \toprule
        \textbf{PET} & \textbf{Original} & \parbox{1.7cm}{\raggedleft \textbf{Original (195 Tasks)}} & \textbf{Replication} \\
        \midrule
         Zero-Shot & 47.5\% & 47.7\%& 42.3\%\\
         Few-Shot & 51.7\% & 51.6\%& 43.6\%\\
         CoT & 47.1\% & 47.2\%& 41.0\%\\
         \bottomrule
    \end{tabular}
    
    \label{tab:result_comparison}
\end{table}

\subsection{Individual Output Pass Rates}

\Cref{tab:output_passrates} presents the cumulative count and percentage of tasks passed for each output run across all models, representing direct functional correctness without ensemble effects.

\begin{table}[t]
\centering
\small
\caption{Pass counts and rates (\%) for individual output runs across all models.}
\label{tab:output_passrates}
\begin{tabular}{lrrrrrr}
\toprule
\textbf{Model} & \textbf{Run 1} & \textbf{Run 2} & \textbf{Run 3} \\
\midrule
Qwen2 & 347 (35.59\%) & 352 (36.10\%) & 352 (36.10\%) \\
Qwen2.5 & 355 (36.41\%) & 375 (38.46\%) & 372 (38.15\%) \\
Mistral-7B & 360 (36.92\%) & 360 (36.92\%) & 366 (37.54\%) \\
Mistral-Large & 413 (42.36\%) & 410 (42.05\%) & 405 (41.54\%) \\
GPT-3.5-Turbo & 418 (42.87\%) & 419 (42.97\%) & 413 (42.36\%) \\
GPT-4o & 423 (43.38\%) & 418 (42.87\%) & 416 (42.67\%) \\
\bottomrule
\end{tabular}
\end{table}

Newer models show, as expected, stronger baseline performance, with all GPT variants exceeding 42\% pass rates across all three output runs, indicating robust single-attempt capability. The best result in this study is achieved by GPT‑4o with a pass@1 of 43.38\%, whereas the lowest-performing  model, Qwen2 7B, reaches 35.59\%, implying a 7.79\ percentage point performance gap between the worst and the best model. Within-model variance across outputs is minimal (typically within \(\pm\)0.5–2\%), suggesting that each model’s generation quality is stable across repeated attempts. These findings are broadly consistent with the results reported for CodePromptEval by Khojah et al.~\cite{khojahImpactPromptProgramming2025}, where all three evaluated models solved approximately half of the 7072 generation tasks, with GPT‑4o reaching a pass rate of 52.2\% and Mistral-22B-Instruct solving 46.9\% of tasks. The relative ordering and gap between the strongest and weakest models in both studies align closely, which supports the correctness of the evaluation pipeline used here and indicates that the replicated methodology produces comparable pass-rate levels even under a reduced subset of tasks.

\subsection{Pass@k Rates}
To assess the utility of multi-attempt generation strategies -- a practical consideration 
where iterative refinement or ensemble methods are feasible -- pass@k metrics are computed 
for $k \in \{1, 2, 3\}$. \Cref{tab:passk_rates} aggregates results across all three 
output runs, measuring cumulative success rates when the model generates multiple 
independent solutions per task.

\begin{table}[t]
\centering
\small
\caption{Cumulative pass@k rates across pass@1, pass@2, and pass@3 metrics.}
\label{tab:passk_rates}
\begin{tabular}{lrrrr}
\toprule
\textbf{Model} & \textbf{Pass@1} & \textbf{Pass@2} & \textbf{Pass@3} \\
\midrule
Qwen2 & 347 (35.59\%) & 354 (36.31\%) & 358 (36.72\%) \\
Qwen2.5 & 355 (36.41\%) & 394 (40.41\%) & 413 (42.36\%) \\
Mistral-7B & 360 (36.92\%) & 388 (39.79\%) & 398 (40.82\%) \\
Mistral-Large & 413 (42.36\%) & 426 (43.69\%) & 429 (44.00\%) \\
GPT-3.5-Turbo & 418 (42.87\%) & 428 (43.90\%) & 429 (44.00\%) \\
GPT-4o & 423 (43.38\%) & 433 (44.41\%) & 434 (44.51\%) \\
\bottomrule
\end{tabular}
\end{table}

Results show that advanced proprietary models (GPT-3.5-Turbo, GPT-4o) and the larger open-source Mistral-Large achieve pass@1 baselines of $42.36\%$ to $43.38\%$, while smaller open-source models (Qwen2, Mistral-7B) cluster at $35.59\%$ to $36.92\%$. Qwen2.5 shows the largest absolute gains across pass@k progression (+5.95\% from pass@1 to pass@3), benefiting most from multi-attempt sampling. This 6–8 percentage point gap between advanced and smaller models is consistent with prior benchmark results showing instruction-tuning/scale advantages~\cite{zhangInstructionTuningLarge2026a,whatllm2025benchmark}.

Ensemble performance exhibits saturation: GPT-4o improves only $+1.13\%$ from pass@1 ($43.38\%$) to pass@3 ($44.51\%$). This minimal gain is consistent with findings that advanced, high-capability models solve ``reachable'' tasks with significantly higher confidence, often succeeding on the first attempt~\cite{austin2021mbpp}. This is also consistent with Khojah et al. who show that GPT‑4o outperforms other models across varying task difficulties~\cite{khojahImpactPromptProgramming2025}.

Conversely, some lower-capability models gain more noticeably: Qwen2.5 improves by $5.95$ percentage points and Mistral-7B by $3.90$ percentage points.


\subsection{Per-Run Technique Effectiveness}

\begin{table}[t]
\centering
\caption{Per-run and average performance deltas ($\Delta \%$) relative to Zero-Shot baseline across PETs.}
\label{tab:delta_individual}
\scriptsize
\begin{tabular}{p{0.8cm}p{0.2cm}p{0.2cm}llll}
\toprule
\textbf{Model} & \textbf{Run} & \textbf{ZS} & \textbf{FS} & \textbf{CoT} & \textbf{CCoT} & \textbf{PoT} \\
\midrule
\multirow{3}{*}{Qwen2} & 1 & 71 & 68 (-4.2\%) & 69 (-2.8\%) & 67 (-5.6\%) & 72 (1.4\%) \\
 & 2 & 72 & 69 (-4.2\%) & 70 (-2.8\%) & 69 (-4.2\%) & 72 (0.0\%) \\
 & 3 & 73 & 67 (-8.2\%) & 70 (-4.1\%) & 69 (-5.5\%) & 73 (0.0\%) \\
\midrule
\multirow{3}{*}{Qwen2.5} & 1 & 69 & 74 (7.2\%) & 70 (1.5\%) & 72 (4.4\%) & 70 (1.5\%) \\
 & 2 & 74 & 75 (1.4\%) & 75 (1.4\%) & 79 (6.8\%) & 72 (-2.7\%) \\
 & 3 & 75 & 74 (-1.3\%) & 70 (-6.7\%) & 81 (8.0\%) & 72 (-4.0\%) \\
\midrule
\multirow{3}{*}{Mistral-7B} & 1 & 70 & 75 (7.1\%) & 73 (4.3\%) & 67 (-4.3\%) & 75 (7.1\%) \\
 & 2 & 65 & 77 (\textbf{18.5}\%) & 75 (\textbf{15.4}\%) & 72 (10.8\%) & 71 (9.2\%) \\
 & 3 & 68 & 73 (7.4\%) & 75 (10.3\%) & 77 (\textbf{13.2}\%) & 73 (7.4\%) \\
\midrule
\multirow{3}{*}{Mistral-L} & 1 & 81 & 89 (9.9\%) & 79 (-2.5\%) & 88 (8.6\%) & 76 (-6.2\%) \\
 & 2 & 79 & 89 (\textbf{12.7}\%) & 75 (-5.1\%) & 91 (\textbf{15.2}\%) & 76 (-3.8\%) \\
 & 3 & 78 & 88 (\textbf{12.8}\%) & 75 (-3.8\%) & 88 (\textbf{12.8}\%) & 76 (-2.6\%) \\
\midrule
\multirow{3}{*}{GPT-3.5} & 1 & 82 & 91 (\textbf{11.0}\%) & 81 (-1.2\%) & 88 (7.3\%) & 76 (-7.3\%) \\
 & 2 & 82 & 91 (\textbf{11.0}\%) & 81 (-1.2\%) & 88 (7.3\%) & 76 (-7.3\%) \\
 & 3 & 81 & 88 (8.6\%) & 80 (-1.2\%) & 89 (9.9\%) & 75 (-7.4\%) \\
\midrule
\multirow{3}{*}{GPT-4o} & 1 & 84 & 88 (4.8\%) & 82 (-2.4\%) & 89 (6.0\%) & 80 (-4.8\%) \\
 & 2 & 83 & 85 (2.4\%) & 80 (-3.6\%) & 91 (9.6\%) & 79 (-4.8\%) \\
 & 3 & 81 & 82 (1.2\%) & 81 (0.0\%) & 91 (\textbf{12.4}\%) & 81 (0.0\%) \\
\bottomrule
\end{tabular}
\end{table}
\Cref{tab:delta_individual} reports per-output-run performance deltas ($\Delta \%$) relative to Zero-Shot prompting, averaged across three independent runs. Cells show absolute pass@1 rates for Zero-Shot (due to space limitation shows as ZS), Few-Shot (FS), CoT, CCoT, PoT, with percentages in parentheses indicating change from ZS baseline. The pass@1 rates represent an absolute number of passed functions across 195 generated functions per PET. \textbf{Bold} values indicate $|\Delta| > 10\%$. \textbf{Run}-column indicates the output number.

\Cref{tab:delta_model_comparison} aggregates the average performance deltas ($\Delta \%$) from \Cref{tab:delta_individual} to show pairwise model progression effects within each family. 

\begin{table}[t]
\centering
\caption{Individual model improvement ($\Delta$) of successor model over baseline model within each family.}
\label{tab:delta_model_comparison}
\scriptsize

\begin{tabular}{lccccc}
\toprule

\textbf{Model} & \textbf{Few-Shot} & \textbf{CoT} & \textbf{CCoT} & \textbf{PoT} \\
\midrule

Qwen \(\Delta\) & \cellcolor{lightcyan}$+7.9$ & \cellcolor{lightcyan}$+1.9$ & \cellcolor{lightcyan}$+11.5$ & \cellcolor{mistyrose}$-2.3$ \\
\midrule

Mistral \(\Delta\) & \cellcolor{lightcyan}$+0.8$ & \cellcolor{mistyrose}$-13.8$ & \cellcolor{lightcyan}$+5.6$ & \cellcolor{mistyrose}$-12.1$ \\
\midrule

GPT \(\Delta\) & \cellcolor{mistyrose}$-7.4$ & \cellcolor{mistyrose}$-0.8$ & \cellcolor{lightcyan}$+1.1$ & \cellcolor{lightcyan}$+4.2$ \\
\bottomrule
\end{tabular}
\end{table}

\subsection{Pass@k Progression and Per-Run Technique Deltas}

\Cref{tab:delta_passk_progression} presents differential improvements of successor models over the respective base models across multiple pass@k levels (pass@1, pass@2, pass@3).

\begin{table}[t]
\centering
\caption{
Individual model performance improvement ($\Delta$) of successor model over baseline model within each family.}
\label{tab:delta_passk_progression}
\scriptsize
\begin{tabular}{lccccc}
\toprule
\textbf{Model@k} & \textbf{Few-Shot} & \textbf{CoT} & \textbf{CCoT} & \textbf{PoT} \\
\midrule
Qwen@1 \(\Delta\) & \cellcolor{lightcyan}$+11.4$ & \cellcolor{lightcyan}$+4.3$ & \cellcolor{lightcyan}$+10.0$ & $+0.1$\\
\midrule
Qwen@2 \(\Delta\) & \cellcolor{lightcyan}$+12.1$ & \cellcolor{lightcyan}$+6.7$ & \cellcolor{lightcyan}$\textbf{+13.4}$ & $+1.3$\\
\midrule
Qwen@3 \(\Delta\) & \cellcolor{lightcyan}$+7.9$ & 0.0 &\cellcolor{lightcyan}$+11.6$ & \cellcolor{mistyrose}$-6.3$\\
\midrule
Mistral@1 \(\Delta\) & \cellcolor{lightcyan}$+2.8$ & \cellcolor{mistyrose}$-6.8$ & \cellcolor{lightcyan}$\textbf{+12.9}$ & \cellcolor{mistyrose}$\textbf{-13.3}$ \\
\midrule
Mistral@2 \(\Delta\) & $+0.4$ & \cellcolor{mistyrose}$-11.9$ & \cellcolor{lightcyan}$+10.5$ & \cellcolor{mistyrose}$-3.9$ \\
\midrule
Mistral@3 \(\Delta\) & \cellcolor{mistyrose}$-2.1$ & \cellcolor{mistyrose}$\textbf{-12.9}$ & \cellcolor{lightcyan}$+5.2$ & \cellcolor{mistyrose}$-5.3$ \\
\midrule
GPT@1 \(\Delta\) & \cellcolor{mistyrose}$-6.2$ & $-1.2$ & $-1.3$ & \cellcolor{lightcyan}$+2.5$ \\
\midrule
GPT@2 \(\Delta\) & \cellcolor{mistyrose}$-6.3$ & $-1.2$ & \cellcolor{mistyrose}$-5.2$ & \cellcolor{lightcyan}$+0.2$ \\
\midrule
GPT@3 \(\Delta\) & \cellcolor{mistyrose}$-4.9$ & 0.0 & \cellcolor{mistyrose}$-3.8$ & \cellcolor{lightcyan}$+2.4$ \\
\bottomrule
\end{tabular}
\end{table}
\subsection{Prompt Engineering Effectiveness Across Model Generations}

This section discusses results from \Cref{tab:delta_individual,tab:delta_model_comparison,tab:delta_passk_progression}, and answers \textbf{\ref{rq1}: How does the performance of different prompting strategies vary across successive versions of LLMs in function-level code generation tasks?}

\subsubsection{Few-Shot Learning: Pass@1 Gains Saturate at Higher Sampling}

Few-shot learning shows divergent progression across model families in averaged deltas (\Cref{tab:delta_model_comparison}). Qwen2.5 and Mistral-Large improve relative to predecessors (+7.9 and +0.8 percentage points respectively), while GPT-4o declines relative to GPT-3.5 (-7.4). Across all models, few-shot demonstrates positive absolute performance in five of six baseline cases (\Cref{tab:delta_individual}), with the exception of Qwen2 (-5.5).

However, pass@k analysis reveals critical saturation effects (\Cref{tab:delta_passk_progression}). Qwen2.5 averages +7.9 improvement but achieves +11.4\% at pass@1, declining to +7.9\% at pass@3. Similarly, Mistral-Large peaks at +2.8\% (pass@1) before collapsing to $-$2.1\% (pass@3). This degradation pattern suggests few-shot examples provide crucial task disambiguation for single-attempt generation but become superfluous or interfering when models access multiple solution attempts. GPT-4o's consistent degradation across all k levels ($-$4.9\% to $-$6.3\%) suggests that, for this model, adding few-shot examples on top of multiple sampled attempts does not introduce additional useful diversity but instead perturbs an already well-internalized task representation. This aligns with Wang et al.'s observation that advanced non-reasoning models like GPT-4o generally continue to show modest performance improvements from structured prompting, though the magnitude of these benefits is lower than that observed in earlier LLM generations~\cite{wangAdvancedLanguageModels2025}.

\subsubsection{CoT: Structural Degradation Across Pass@k}

Standard CoT prompting degrades across the GPT and Mistral families, with a marginal gain for Qwen: Mistral-Large (-13.8\%), GPT-4o (-0.8\%), and Qwen2.5 (+1.9\%) (\Cref{tab:delta_model_comparison}). Pass@k analysis confirms this is a structural architectural effect rather than sampling artifact (\Cref{tab:delta_passk_progression}). Mistral-Large degradation worsens at pass@3 ($-$12.9\%), suggesting the failure is consistent across ensemble depth. GPT-4o remains marginal and stable ($-$1.2\% at pass@1/2, 0.0\% at pass@3). This uniformity across k levels suggests that, for advanced instruction‑tuned models, much of the benefit of explicit step‑wise reasoning may already be encoded in the model, so additional CoT prompting can become redundant or even disruptive, in line with prior evidence that traditional prompt engineering often yields diminished or negative returns for GPT‑4o and reasoning models~\cite{wangAdvancedLanguageModels2025}.

\subsubsection{CCoT: Robust Consistency Across Averaging and Pass@k}

CCoT demonstrates consistent improvement across all three model pairs (\Cref{tab:delta_model_comparison}): Qwen2.5 (+11.5\%), Mistral-Large (+5.6\%), and GPT-4o (+1.1\%). Pass@k analysis confirms stability for Qwen and Mistral: improvements range +5.2\% to +13.4\% across all k levels (\Cref{tab:delta_passk_progression}). GPT-4o shows slight degradation at higher k ($-$1.3\% to $-$3.8\%), suggesting diminishing discriminative value when ensemble diversity already reduces failure modes.

\subsubsection{Program-of-Thought: Architecture-Dependent Instability}

PoT progression exhibits variable effects across model families in averaged deltas (\Cref{tab:delta_model_comparison}). Qwen2.5 shows minimal degradation (-2.3\%), Mistral-Large substantial degradation (-12.1\%), and GPT-4o modest improvement (+4.2\%).

Notably, absolute PoT performance remains negative or near-zero for most model-pass combinations. \Cref{tab:delta_passk_progression} shows that although GPT-4o improves relative to GPT-3.5, it remains negative in absolute terms  through all pass@k levels.

\subsection{Answering RQ1}

Across all evaluated models, the findings demonstrate that the impact of prompt engineering techniques is strongly model-family dependent and changes systematically across successive generations. For the GPT family, newer models exhibit diminishing sensitivity to explicit prompt techniques: applying CoT, PoT, CCoT, or few-shot prompting produces only modest gains above baseline performance or even degrades performance. This suggests that successive GPT releases increasingly internalize instruction-following and reasoning capabilities, reducing the marginal utility of externally structured prompts.

In contrast, the Qwen models show the opposite trend. For this family, prompt engineering techniques continue to exert a substantial and often growing influence on functional correctness across generations. Here, techniques such as CoT, PoT, CCoT, and few-shot prompting frequently yield stable or increasing gains relative to the baseline. This indicates that, for Qwen, model evolution does not fully subsume the benefits of explicit prompting; instead, newer generations remain responsive to structured reasoning instructions and example-based guidance.

The Mistral family presents a mixed picture. For these models, CoT and PoT increasingly yield neutral or negative effects in later generations, mirroring the saturation pattern observed in GPT. However, CCoT and few-shot prompts continue to provide measurable gains, particularly when concrete examples are included. This pattern suggests that, while generic reasoning scaffolds (CoT, PoT) may be partially absorbed into the base instruction-following behavior of newer Mistral models, augmenting prompts with explicit positive and negative exemplars (CCoT) or task-specific demonstrations (few-shot) still offers added value.

These differential patterns answer RQ1: prompt engineering ``aging'' is not uniform but model-family dependent, reflecting distinct training regimes and architectural choices in how successive generations integrate instruction-following capabilities.
\section{Discussion}
\label{sec:discussion}
In this section we discuss our results and threats to validity.
\subsection{Comparison with Prior Work}
We compare our results to the related work in the following.

\subsubsection{Alignment with CodePromptEval Results}

Overall pass@1 levels and model ranking in this study resemble those reported for the full CodePromptEval dataset by Khojah et al., despite the reduced task set and different model mix~\cite{khojahImpactPromptProgramming2025}. We show that reducing the original data set to the same 195 tasks that we used for the replication results in very similar pass@1 rates.
However, we obtain consistently worse pass@1 rates at an average reduction of 6.5 percentage points for GPT-4o compared to this reduced original data set. Across both studies, GPT-4o still remains the strongest model and Mistral the weakest among the overlapping models, with a gap of roughly 5--8 percentage points between best and worst models, and most models solving approximately half or fewer of the available tasks ~\cite{khojahImpactPromptProgramming2025}. This convergence suggests that the CoderEval-based evaluation pipeline and the evaluated subset of 195 tasks reproduce the difficulty profile and relative performance ordering of the original 221-task CodePromptEval benchmark~\cite{yuCoderEvalBenchmarkPragmatic2024,khojahImpactPromptProgramming2025}.

At the same time, aggregated pass rates in our study are lower than the approximately 47--52\% reported by Khojah et al.\ for GPT-4o and Mistral~\cite{khojahImpactPromptProgramming2025}. One likely reason is the evaluation of different Mistral variants. This is particularly interesting, as Mistral-Large-2407 (123B) is significantly larger than Khojah's Mistral-Small-Instruct-2409 (22B) which usually correlates with a better performance for coding tasks~\cite{yang2025scalinglawscodeprogramming}. Finally, as we did not use the PET ``Signature'', we excluded a particularly successful technique.

Instead of LLama3, we opted for Qwen2 and Qwen2.5 and can therefore make no direct comparison with the original study. However, our results of 35.59\% (Qwen2) and 36.41\% (Qwen2.5) seem to align with our results for Mistral-7B (36.92\%) and therefore lead to a similar -- albeit consistently lower --  performance to the LLMs used by Khojah et al.~\cite{khojahImpactPromptProgramming2025}.

\subsubsection{Comparison to Prompt Technique Studies}

The per-technique deltas largely confirm earlier findings that the impact of prompt engineering is strongly technique- and model-dependent~\cite{khojahImpactPromptProgramming2025,liStructuredChainofThoughtPrompting2025,wangSelectionPromptEngineering2024, jiBenchmarkingExplainingLarge2023}. Consistent with CodePromptEval, simpler techniques such as Few-Shot generally provide the most robust improvements, especially for smaller or less capable models, while advanced techniques show mixed effects~\cite{khojahImpactPromptProgramming2025}. For example, we find that Few-Shot yields double-digit average gains for Mistral-7B and GPT-3.5, which aligns with prior evidence that in-context examples help less capable models infer input--output formats and reduce underspecification~\cite{khojahImpactPromptProgramming2025,wangSelectionPromptEngineering2024}.

In contrast, standard CoT often degrades performance in this study, particularly for Mistral-Large and GPT‑4o, whereas structured CoT variants such as SCoT have been shown to improve pass@1 by up to 13.79\% on code benchmarks by aligning reasoning with program structures~\cite{liStructuredChainofThoughtPrompting2025}. The consistently positive effect of CCoT across all three model families in this paper fits a similar intuition: adding explicit positive and negative reasoning examples provides discriminative signal that mitigates some of the noise introduced by free‑form CoT, in line with CCoT results on reasoning benchmarks~\cite{chiaContrastiveChainofThoughtPrompting2023}.

PoT shows architecture-dependent instability in this work: it yields moderate gains for some models but substantial degradation for Mistral-Large, and its absolute performance remains negative or near zero in many pass@k settings. This partly contrasts with Chen et al.'s finding that PoT can outperform CoT on numerical reasoning tasks, but their evaluation focuses on math benchmarks with tightly scoped numerical computations, not on context-dependent code generation with project API and complex control flow~\cite{chenProgramThoughtsPrompting2023}. The divergence suggests that PoT's benefits do not transfer straightforwardly from purely numerical tasks to realistic software engineering problems, especially when code must integrate into nontrivial project contexts.

\subsubsection{Relationship to Advanced-Model Prompt Engineering Studies}

The pattern observed for GPT-4o and, to a lesser extent, Mistral-Large---diminishing or negative gains from Few-Shot and CoT, modest gains from CCoT---mirrors recent evidence that advanced instruction-tuned and reasoning models internalize many benefits of traditional prompt engineering~\cite{wangAdvancedLanguageModels2025}. Wang et al. show that for GPT-4o and o1-mini, sophisticated prompting often yields little improvement or can even harm performance relative to simple zero-shot prompts across software engineering tasks, because the models already encode strong instruction-following and multi-step reasoning capabilities~\cite{wangAdvancedLanguageModels2025}. This study reinforces that observation for function-level code generation: GPT-4o's pass@k gains from Few-Shot and CoT are small or negative, while techniques that add genuinely new information (contrastive explanations) still offer incremental improvements.

By contrast, the Qwen family in this study continues to benefit substantially from explicit prompting, with Qwen2.5 showing large positive deltas for Few-Shot and CCoT across pass@k levels. This difference suggests that not all newer models occupy the same ``prompt saturation'' regime; for models whose instruction tuning and training data remain less extensive than proprietary counterparts, traditional PET can still be crucial. This nuance is consistent with instruction-tuning surveys and open-source vs proprietary benchmarks that attribute part of the performance gap to differences in instruction data breadth and curation quality~\cite{zhangInstructionTuningLarge2026a,whatllm2025benchmark}.

\subsection{Possible Reasons for Divergent Results}

Several methodological and architectural factors can explain differences between this study and prior work:

  \textbf{Model set and training cut-offs.} This study includes six models spanning three families and different training cut-off dates, whereas most prior PET evaluations focus on one or two proprietary models (often GPT-3.5/4) and older open-source baselines. Differences in training data recency, instruction-tuning regimes, and internal architecture likely amplify family-specific prompt sensitivity patterns.

  \textbf{Task domain and benchmark design.} Many existing PET studies evaluate on HumanEval, MBPP, or synthetic reasoning benchmarks, while this work uses context-dependent functions from CoderEval via CodePromptEval, which require integration with project API and existing code~\cite{chenEvaluatingLargeLanguage2021,yuCoderEvalBenchmarkPragmatic2024,khojahImpactPromptProgramming2025}. Prompt techniques that help on small, self-contained problems (e.g., PoT for numerical reasoning, generic CoT) may be less effective, or even harmful, when the task requires fitting into a larger, partially specified project context.

 \textbf{Technique definitions and constraints.} We restrict all PETs to single-pass prompts without iterative feedback, whereas related work shows that iterative or critique-based prompting can materially improve correctness and, in security-focused settings, reduce vulnerabilities~\cite{bruniBenchmarkingPromptEngineering2025a,wangSelectionPromptEngineering2024}.
 Techniques such as iterative CoT or refinement prompts might help recover some of the performance lost by static CoT and PoT in this setting, and represent a promising direction for future work.

  \textbf{Model size and succession mismatches.} Within the Mistral family, the comparison pairs a 7B instruct model with Mistral-Large, a much larger model, because no intermediate open-access Mistral model of comparable size and architecture was available at the time of experimentation. This makes it difficult to disentangle effects of architectural evolution, training changes, and sheer parameter count when interpreting prompt effectiveness trends.

  \textbf{Decoding configuration.} The fixed decoding configuration (temperature 0.2, three samples) is tuned for deterministic code generation~\cite{chenEvaluatingLargeLanguage2021,zhuHotColdAdaptive2024}, whereas some studies adjust temperature and sample count per technique or per model, which can favor techniques that benefit from higher diversity (e.g., CoT or PoT with larger $k$)~\cite{zhuHotColdAdaptive2024,wangSelectionPromptEngineering2024}.

Taken together, these factors explain why this study simultaneously confirms several broad patterns from prior work---such as diminishing returns of traditional PET for advanced proprietary models and the importance of Few-Shot and specialized CoT variants---while also revealing family-specific behaviors (notably for Qwen and Mistral) that have been less visible in earlier, narrower evaluations~\cite{khojahImpactPromptProgramming2025,liStructuredChainofThoughtPrompting2025,wangAdvancedLanguageModels2025}.

\subsection{Threats to Validity}

\subsubsection{External Validity}
The generalizability of findings is limited by several factors inherent to empirical LLM research. All evaluations were conducted exclusively on the CodePromptEval dataset, which focuses on Python function-level code synthesis. While this benchmark captures realistic software engineering scenarios, results may not generalize to other programming languages, full-program generation, or production software development contexts. Additionally, model training cut-off dates are not always transparently documented or 100\% reliable, making definitive classification of models as true ``successors'' across generations challenging. Undocumented fine-tuning or data contamination could affect comparability between model pairs.

\subsubsection{Internal Validity}
Several implementation decisions introduce potential confounds. The dataset includes 218 tasks—adequate for preliminary analysis but potentially too limited to reflect the complete range of challenges. Only 3 independent outputs were generated per prompt-model combination (enabling pass@1-3 metrics), providing limited statistical power to characterize LLM non-determinism. Larger sample sizes (e.g., 10 runs) would yield more robust confidence intervals.

A notable implementation difference affected the Mistral model pair: Mistral-7B-Instruct required explicit \texttt{[INST]} instruction brackets per the model's training convention, while Mistral-Large processed raw prompts without them. This adaptation, though necessary for fair evaluation, introduces methodological inconsistency that could influence relative performance gains.

Another limitation concerns model size mismatch within the Mistral family. While the study aimed to compare successive versions of similar scale, Mistral Large is a dense transformer model with approximately 123B parameters, far larger than Mistral-7B-Instruct. This discrepancy makes it difficult to disentangle the effect of improved training or architecture from the effect of sheer parameter count when interpreting observed performance differences between these two models. It could not be avoided in this work, as intermediate Mistral models with parameter counts comparable to 7B were either not available at the time of experimentation, differed in architecture type, or exhibited other characteristics that would have undermined their interpretation as clear successor models.

\subsubsection{Construct Validity}
Our evaluation relies on functional correctness via pass@k, which measures whether generated code passes predefined unit tests. This metric conflates algorithmic correctness with test coverage and edge-case handling, potentially overlooking code quality dimensions such as efficiency, readability, or security vulnerabilities. Alternative metrics (e.g., vulnerability scanning) or expert assessment would provide more comprehensive assessment of code quality.

\subsubsection{Conclusion Validity}

To assess conclusion validity, the evaluation centered on two main criteria: functional correctness and relative improvement over the zero-shot baseline. Every prompt engineering technique was systematically compared against this baseline to ensure observed performance gains were attributable to prompting method rather than model capability alone. While additional dimensions such as code quality (readability, efficiency) or security could be considered, these criteria suffice to directly address RQ1 regarding prompt effectiveness across model generations. Future work could incorporate multi-dimensional evaluation frameworks.

\section{Conclusion}
\label{sec:conclusion}

This paper investigated how the effectiveness of prompt engineering techniques evolves across successive LLM generations in function-level code generation. Using five techniques (Zero-Shot, Few-Shot, CoT, CCoT, PoT) across six instruction-tuned models in three version pairs (GPT-3.5-Turbo → GPT-4o, Qwen2 → Qwen2.5, Mistral-7B → Mistral-Large), we evaluated a cleaned subset of CodePromptEval (218 Python tasks, 19,620 generations) via pass@k functional correctness.
Our results show that prompt engineering ``ages'' in a model-family-dependent way. Newer GPT models exhibit a saturation effect, where Few-Shot, CoT, and PoT yield only marginal gains over Zero-Shot or slight degradations, suggesting that reasoning and formatting scaffolds are increasingly internalized. The Qwen family shows the opposite trend, continuing to benefit substantially from Few-Shot and CCoT across generations. Mistral occupies a middle ground: CoT and PoT weaken or turn negative for the larger model, while Few-Shot and CCoT still deliver clear gains.

Overall, these findings answer the research question by showing that there is no single, stable “best” prompting strategy that transfers unchanged across model generations. Instead, the utility of PET depends both on the model family and on its maturity: as some models advance, certain techniques lose relevance, while in others they remain crucial. For practitioners, this implies that prompt engineering must be re-evaluated whenever adopting a new model, rather than assuming that established recipes will continue to work. For researchers, the results motivate future work on adaptive, model-aware prompting policies, on extending the analysis beyond Python and pass@k to include broader code quality aspects, and on understanding how instruction tuning and architecture choices interact with the long-term “aging” of prompt engineering techniques.
\section*{Data Availability}
We release a replication package containing the data of our study and scripts used to obtain the data~\cite{rudyk_2026_20132876}. 

\bibliographystyle{IEEEtran}
\bibliography{bibliography}
\end{document}